\documentclass[aps,prl,twocolumn,groupedaddress,showpacs]{revtex4}

\usepackage{graphicx}
\usepackage{amssymb}

\def\he4{$^4$He}
\def\hee3{$^3$He}
\def\t0{$T_0$}

\begin{document}

\title{Possible Vortex Fluid  to  Supersolid Transition  in Solid \he4 below $\sim$75 mK}
\author{Nobutaka Shimizu, Yoshinori Yasuta,  and Minoru Kubota}
\email{kubota@issp.u-tokyo.ac.jp}

\affiliation{Institute for solid State Physics, University of
Tokyo, Kashiwanoha 5-1-5, Kashiwa, 277-8581, Japan }

\date{\today}

\begin{abstract}
A detailed torsional oscillator(TO) study on  a  stable  solid \he4 sample at 49 bar with  $T_o$$\sim$0.5K,  is reported to $T$ below the dissipation peak at $T_p$. 
We  find both the shift of period and dissipation hysteretic  behavior 
below $T_c$ $\sim$75 mK, with changes of AC excitation amplitude $V_{ac}$.  The derived difference of non-linear rotational susceptibility $\Delta$$NLRS(T)_{hys}$  across the hysteresis loop under systematic conditions  is analyzed as a function of $V_{ac}$ and $T$. 
We propose that $\Delta$$NLRS(T)_{hys}$ is the 
non-classical rotational inertia fraction, $NCRIF$ itself, and it 
is actually the supersolid density $\rho_{ss}$ of the 3D supersolid state below   $T_c$.  $\rho_{ss}$ changes linearly with $T$ down to$\sim$60 mK and then increases much more steeply, approaching a finite value towards $T$=0. 
We  find a characteristic AC velocity $\sim$40$\mu$m/s beyond which the hysteresis  starts
at $T< T_c$ and a "critical AC velocity"$V_c$$\sim$10 mm/s, above which $\rho_{ss}$ is completely destroyed. We  
 obtain  $\xi_0$ and $V_c$=h/($m_4$$\centerdot$$\xi_0\centerdot\pi$)=$\sim$6-12 mm/s. 
\end{abstract}
\maketitle

The supersolid state, which can be characterized as a solid 
with a lattice structure yet simultaneously having superfluid properties, has been one of the most interesting topics in condensed matter physics. 
Discussions started in the 1960's (see recent reviews\cite{prok, ssdisorder}) and the experimental search for such a state in quantum solids followed for 
40 years till now. Unusual properties of solid \he4 
 have been reported for some time. The first convincing claim was made by Kim and Chan\cite{kimchan}, in a report of 
non-classical rotational inertia ($NCRI$), a phenomenon predicted long ago  by 
 Leggett\cite{leggett}. This 
 experimental observation with a torsional oscillator(TO) has been 
confirmed by other groups including the present authors' group\cite{shirahama, rittnerreppy,
rittnerreppy1,ourqfs, vfluid}. There were, however, a few fundamental problems in the identification of  $NCRI$ in the 
reported experimental results by that time. The problems were discovered following newer observations, namely, the reported onset temperature $T_o$ of 0.1- 0.5 K\cite{vfluid} is too high 
for the known number density of imperfections in solid \he4 samples\cite{imperfection} to expect Bose-Einstein Condensation (BEC) on one hand, while the reported amount of the $NCRI$ fraction goes up to $\sim$ 20 \% of the total  \he4 mass\cite{rittnerreppy1} in conflict with Leggett's original ideas about the origin of supersolidity as 
BEC in the imperfections. 
Furthermore,  the failure to observe superflow\cite{DCflow2005, DCflow2006}
 is suggestive of some other phenomenon, 
which is actually responsible. 
Anderson proposed a picture of a vortex fluid(VF) state above a real $T_c$\cite{andersonvortexfluid, andersonVF2008} for the reported observations for the solid \he4. He took the 
absence of superflow into account 
and the sensitivity to the amount of disorder\cite{rittnerreppy1} and proposed the 
reported TO response is the "nonlinear rotational susceptibility", $NLRS$ of the VF state, instead of $NCRI$ for the real supersolid state and the latter should exist at some lower temperature below $T_c$. Actually another independent picture had been proposed by Shevchenko long ago\cite{shev87, shev88}, considering a 1D dislocation core as an origin of superfluidity. It describes a real superfluid state in solid He below some $T_c$ and still they expect something like superfluid response above $T_c$ because of the dynamic properties. We do not consider 
this picture 
now because we do not find a direct connection with 
 the experimentally found phase transition. 

It is well-known that the VF  is a state without superflow, or  without 3D macroscopic coherence in the case of underdoped(UD) cuprates\cite{wang}. It is also argued to be characterized by  almost constant amplitude of the macroscopic wave function through the real 3D macroscopic super-conducting(-fluid) transition at $T_c$, but phase fluctuations break down the macroscopic coherence\cite{andersonvortexfluid,vringSF} at 
 $T_c$, which can be at much lower $T$ 
 than the onset temperature, $T_o$, where quantized vortices 
 (probably in a low D subsystem)  start to appear thermally.  
As a result of increased low D coherence length $\xi_D$ (which increases towards $T$= 0K), a real 3D supersolid transition should occur at some low $T$. 
A sharp peak in specific heat, which would indicate a 3D real phase transition, has been reported around 75 mK\cite{Cpeak}.

On the other hand, there also have been some attempts to explain features of observations in terms of classical dislocation motion trapped by  \hee3  in connection to the response of solid  \he4 to shear motion apart from supersolid properties\cite{shear}. Actually, a similar shear modulus increase below about 200 mK has been observed not only in hcp solid \he4, but also in hcp solid $^3$He. But a TO response anomaly, $NLRS$ or $NCRI$ has been observed only in hcp \he4\cite{JwestSS2008}.

In a recent publication we have shown\cite{vfluid}  experimental evidence which supports 
the VF picture in solid \he4 samples at 32 bar as well as 49 bar below a common 
$T_o$ $\sim 0.5$ K.  $T_o$ was determined for the first time for solid \he4 from  a detailed study of  the AC excitation $V_{ac}$ dependence change at this temperature\cite{vfluid}. We argued for 
$V_{ac}$ and $T$ dependent responses of 
the pre-existing thermal fluctuations of the phase of a mesoscopic scale wave function in the  VF state. Stronger AC excitations, which would cause formation of  straight vortex lines,  could suppress these thermal fluctuations of the VF state. Such suppression appears not only in 
$NLRS$, but also in the energy dissipation\cite{vfluid}. The unique $log(V_{ac})$ linearly dependent suppression of the $NLRS$ and its unique
temperature dependence suggest that what we observed was not 
$NCRI$, expected for the 3D supersolid, but what is being proposed by Anderson\cite{BoseFluidsAboveTc} as $NLRS$ of the VF state. All the observations, especially above the energy dissipation peak, can be well described  by the properties of the VF state\cite{vfluid, andersonvortexfluid, andersonVF2008}. The VF state may have some  features in common with superfluid turbulence, where the "polarization" of the vortex tangle under rotation may be described as an ensemble of vortex loops and its "polarizability" under 
rotation could show similar behavior to that represented in the Langevin function\cite{Langevin}.
 Similar suppression of fluctuations by an external magnetic field has been reported for the VF state in layered superconductors\cite{layersc}, and cuprate high T$_c$ superconductors\cite{wang}. 
 3D superfluidity and 3D vortex lines are realized in a series of 3D connected He monolayer systems, where 3D connectivity of the superfluid is provided by  the 3D connected surface of a porous substrate\cite{fukuda2005+}. The critical velocity of this system  for  destruction of 
 the 3D superfluidity seems to be characterized by $V_c$$\sim$ $h/m$$a$$\pi$, above which 3D superfluidity is destroyed and 2D superfluid features appear\cite{vc1}, where  $a$ is the 
 3D vortex core size, or the minimum size of the 3D superfluid.
\begin{figure}
\includegraphics[width=0.84\linewidth]
{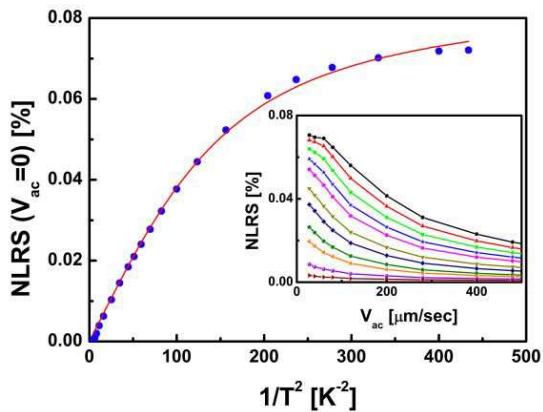}\caption{\label{langevin}$NLRS(T)$ at $V_{ac}$ $\to 0$, is displayed as a function of $1/T^2$. The solid line through the data points is the Langevin function f(x)=a[$\{$exp(bx)+exp(-bx)$\}$/$\{$exp(bx)-exp(-bx)$\}$ - 1/(bx)] with a = 0.0878 $\pm$ 0.0011, and b = 0.0148 $\pm$ 0.0004. Inset shows the  $V_{ac}$ dependence for data at each $T$$\leqq$300 mK and we can safely extrapolate to $V_{ac}$ $\to$ 0.}
\end{figure}
\begin{figure}[t]
\includegraphics[width=0.57\linewidth]
{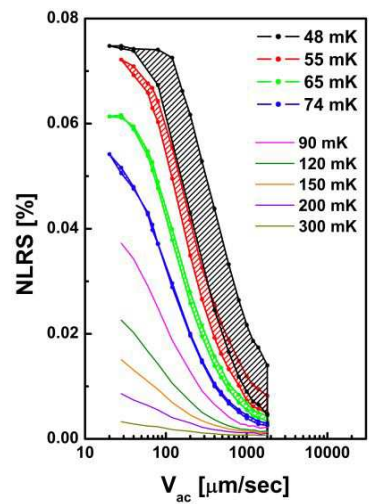}\caption{\label{logvac} The $logV_{ac}$ dependence of $NLRS$ of solid $^4$$He$ sample at 49 bar  at constant $T's$ as given in the figure, obtained from the measurement of period change of TO. 
} 
\end{figure}
%

In the present paper we describe further a detailed TO 
study on the same stable  solid  \he4 sample at 49 bar
\cite{vfluid}, but extended to $T$ lower  than the dissipation peak at $T_p$. The first observation at the lower $T$ is that 
the $logVac$ linear slope's $T$ dependence deviates  
 from the reported $1/T^2$  dependence at higher $T$ for the VF state \cite{vfluid}. 
 To make the physical meaning clearer we consider the value of $NLRS$ extrapolated to $V_{ac}$=0. It would represent the full amount of $NLRS$ of the VF state. Fig.~\ref{langevin} indicates $NLRS(T)$ for  $V_{ac}$ extrapolated to 0 as a function of $1/T^2$. It also deviates from the $1/T^2$ linear dependence at lower $T$ and was found to follow the Langevin function quite well with $x= 1/T^2$ and it approaches a finite value towards $T$=0, ~0.088$\%$ of the total solid $^4He$ mass. All the data in Fig. 1 inset were measured under a certain procedure, where the $V_{ac}$ is set to a maximum 
 value at $T>\sim$0.5K  and  cooled down to desired $T$ and $V_{ac}$ was swept downwards stepwise after equilibrium at each $V_{ac}$. After completion of the measurement at a $T$ another set is repeated for another $T$.  
 We call this procedure "measurement under equilibrium condition".  %
The $T$ dependence may indicate that the VF state is best described by behavior of an ensemble of vortex loops as was discussed for superfluid turbulence in\cite{Langevin}. Loops can be polarized in a manner  similar  to dipole systems, where instead of $1/T^2$,  $1/T$ appears. 
%

In this lower $T$ range where deviation from the $1/T^2$ linear dependence occurs, we find a new feature, namely hysteretic behavior, when an appropriate  
procedure is followed. It appears below a certain temperature $T_c$. 
We will propose that we have observed a transition 
from a vortex fluid state\cite{vfluid} to a new state occurring below 
$T_c$, which is characterized by the appearance of the hysteretic behavior and  of  a critical velocity of the order $\sim$ 1 cm/s, beyond which the hysteretic component of NLRS is suppressed to zero. 
Actually hysteretic behavior itself has been reported by Kojima and his group at very low temperatures\cite{kojima} and  by Chan's group\cite{hystory}, as well as by Reppy's group\cite{reppy08}; however, 
none of them discussed the hysteresis in connection to the transition from the vortex fluid(VF) state. The VF state has been recently experimentally clarified by our report of  the unique $V_{ac}$ and $T$ dependent  TO responses\cite{vfluid}, effects which are absent in known superfluid transition responses.

We describe here the first systematic study of the hysteretic component of the solid \he4 TO responses as a function of $V_{ac}$  as well as $T$ below  $T_c$ and discuss a possible order parameter of the supersolid(SS) state below $T_c$.
\begin{figure}[t]
 \includegraphics[width=0.70\linewidth]
{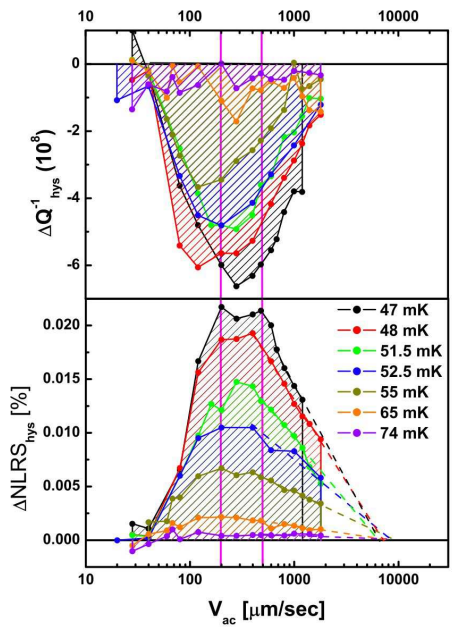}
 \caption{\label{hysVac} "hysteretic"$\Delta$$NLRS_{hys}$= $NCRIF$ as well as $\Delta$$Q^{-1}_{hys}$ 
  vs $log$$V_{ac}$. From the linear extension of the $log$$V_{ac}$ dependence we obtain a critical velocity, 6$\sim$10 mm/s  to suppress the  
 $NCRIF$=$\rho_{ss}$ to zero, which compares well with $V_c$=h/($m_4$$\centerdot$$\xi_0\centerdot\pi$)=$\sim$6-12 mm/s, for $\xi_0$=50-25 nm, see text.}
 \end{figure}
%
%
In Fig.~\ref{logvac} the TO period change as a function of $V_{ac}$ is shown for various $T$. 
 All the data points for each $T$ down to 80 mK follow a single line as was the case for our data in a previous publication 
above $T_p$\cite{vfluid}. The  data points below $~75$ mK actually follow two lines. 
These two lines were produced as follows:  The first measurements were performed at 
 $V_{ac}$=maximum set at $T$ above $500$ mK, as "equilibrium" condition measurements. Then they were prepared  by cooling down the sample  to the lowest $T$$\sim48
 $ mK, and then warmed up to the desired $T<T_c$ and measurements were performed by changing  $V_{ac}$ step wise downwards over a long enough time,  typically 12 hours or longer for the whole sweep one way, to allow for any relaxation  at this constant $T$. 
Then the other series of measurements were prepared  by reversing the excitation $V_{ac}$ change upwards to the measuring excitation step-wise after equilibrium is reached at each step up to the maximum $V_{ac}$. After completing measurements at a $T$ then another set of measurements at a different $T$ was performed in a similar manner. We observe in  Fig.~\ref{logvac}  that the hysteretic component(the difference between the two passages of $V_{ac}$) 
has a unique $V_{ac}$ dependence  and increases towards lower $T$.

\begin{figure}
 \includegraphics[width=0.7\linewidth]
 {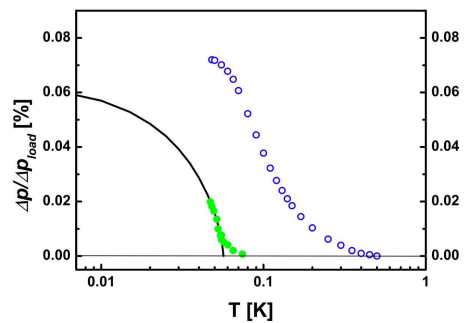}
 \caption{\label{Sponta}The $T$ dependence of the $NCRIF$ =$\rho_{ss}$(at 200 $\mu$m/s)
 (closed symbols) appeared as a result of the "hysteretic process" in \% of the total He mass. It  represents the  "persistent"  $NCRIF$ produced by the process.  
 It would be evidence of 3D macroscopic phase appearance below $T_c$. It compares with "paramagnetic" behavior of $NLRS$($V_{ac}$=0, open symbols), 
 which follows the Langevin function as in Fig. 1,  and suggests a "susceptibility" feature of $NLRS$.} 
\end{figure}
Fig.~\ref{hysVac} shows much more clearly the hysteretic components of both dissipation and $NLRS$ as a function of 
 $V_{ac}$ presented on a logarithmic horizontal scale. An interesting observation is that the hysteretic component appears only above a characteristic AC velocity$\sim40 \mu$m/s and it reaches a maximum at $\sim200 \mu$m/s. Furthermore, at higher AC velocity,$\sim500\mu$m/s it starts to decrease. It looks as if it follows a $log$$V_{ac}$ linear relation  passing through $\Delta$$NLRS_{hys}$= 0 in the range of $V_{ac}$, 8$\pm 2$ mm/s, signaling the total  depression of the hysteretic $NLRS$ component.
The upper column of Fig.~\ref{hysVac} shows the energy dissipation  change across the hysteresis loop $\Delta$$Q^{-1}_{hys}$ as $V_{ac}$. A most impressive thing is that while $\Delta$$Q^{-1}_{hys}$ is negative $\Delta$$NLRS_{hys}$ is  increasing over a considerable range of $V_{ac}$. We suggest this is  evidence for a shielding 
current preventing penetration of vortex lines as well as 
thermally excited vortices into the volume as in the Landau state of superfluids or Meissner state of superconductors. 
At higher AC velocity above $
500\mu$m/s introduction of vortex lines would cause a $log$$V_{ac}$ linear decrease of supersolid density $\rho_{ss}$.

Fig.~\ref{Sponta} indicates the $T$ dependence of the hysteretic component of  
$\Delta$$NLRS_{hys}$($200\mu$m/s) together with "equilibrium" $NLRS$($V_{ac}$$\to$0). The former appears below about 75 mK and increases almost linearly toward lower $T$ to $\sim$ 60 mK, then further increases more steeply as $T$ lowers. We have tried to fit it with the expected behavior for a 3D supersolid density, $\rho_{ss}$$=(1-T/T_c)$$^\gamma$=$t^\gamma$ with $\gamma$=2/3 and it yields 
an extrapolation to 0 K(t=1).  We can  safely  estimate " the supersolid density extrapolated to 0 K" within some  error bar. It would be 0.065$\pm0.025 \%$ of the total solid He sample mass with $T_c$ =56.7 mK. The $\rho_{ss}(T=0)$ is less than but on the same order as the expectation value of $NLRS(T=0)$, which increases towards 0 K following a Langevin function, as discussed.
%
%
\begin{figure}
 \includegraphics[width=0.75\linewidth, ]
 {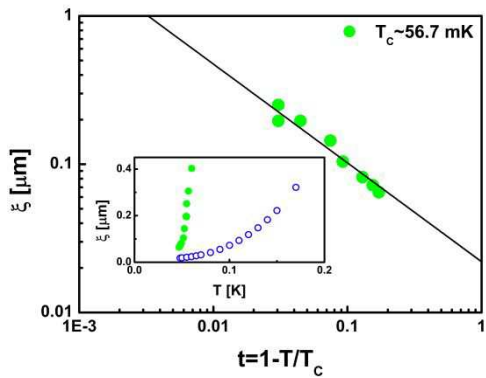}
 \caption{\label{rhos-t}Critical behavior of $\xi$ obtained from the data in Fig.4. 
 We took $T_c$ = 56.7 mK, supposing $\rho_{ss}$$=(1-T/T_c)$$^\gamma$=$t^\gamma$ with $\gamma$=2/3. And we  obtain $\xi_0$$\sim$ 25 to 50 nm 
 by simple extrapolations, horizontal and straight extension of the linear relation.}
 \end{figure}
 %
 %
%
%
From the absolute value of the $\rho_{ss}$, we can evaluate supersolid  coherence length $\xi$, 
following a common practice 
based on consideration of 
Josephson's length \cite{guzai}. We could fit  rather well with the
 $t^{-2/3}$ dependence expected for a 3D superfluid, obtaining an extrapolated $\xi$ value at $T$=0, $\xi_0$=25 to 50nm as shown in Fig.~\ref{rhos-t}, assuming $m_4$ to be the atomic $^4$He mass.  We find  $V_c$=h/($m_4$$\centerdot$$\xi_0$$\centerdot\pi$)=$\sim$6-12 mm/s, for $\xi_0$=50-25 nm, a surprising coincidence.
 It is very  interesting considering the microscopic origin of the 3D macroscopic supersolid phenomenon as suggested by the present work. Actually there have been various observations of the size of $NLRS$, but nobody else discussed the transition from the VF to the SS state except the present authors. 
%
%
%
%
 %
%
%
%
%
A similar hysteresis phenomenon was already experimentally observed for a $^4$He sample with 30 ppm $^3$He long  ago in an acoustic experiment by Iwasa and Suzuki in 1980\cite{iwasahys}, but without noticing the 
relationship to a vortex state. 

We have found  the start of the hysteretic behavior below $T_c$, and evaluated the hysteretic component $\Delta$$NLRS_{hys}$ as a function of $V_{ac}$ and discussed it as supersolid density  $\rho_{ss}(T)$ and discussed the 3D coherence length $\xi$ and a consistent critical AC velocity $V_c$$\sim$1 cm/s. So far we have neglected the anisotropy of the hcp crystal of $^4He$ and treated it as an isotropic supersolid. The linear $T$ dependence of $\rho_{ss}(T)$ between 75 and $\sim$ 60 mK may have some connection to this problem.
%

\begin{acknowledgments}
The authors  acknowledge A. Penzev's  work at the initial stage, and  T. Igarashi, P. Gumann, T. Miya and R.M. Mueller's assistance. M.K. is thankful for valuable discussions with P.W. Anderson, D. Huse, W. Brinkman, J. Beamish and many other colleagues in a series of workshops organized by M. Chan and D. Ceperley, by Shirahama, by Prokof'ev, Svistunov and  Stamp, by Sonin, by Jezowskii, as well as by Balibar, M. Chan, Kiselev and Svistunov. I. Iwasa helped us understand dislocations studies in solid He. We thank him also for pointing out of their early work on hysteresis 
 in \cite{iwasahys}.

\end{acknowledgments}

\bibliography{vftoss}
\end{document}